\documentclass{eas}
\usepackage{graphicx}
\begin{document}

\title{From Reference Frames to Relativistic Experiments:
Absolute and Relative Radio Astrometry} 
\author{E.~B.~Fomalont}\address{National Radio Astronomy Observatory, Charlottesville, VA 22903, USA}
%
%
\begin{abstract}

Reference systems and frames are crucial for high precision {\it
absolute} astrometric work, and their foundations must be
well-defined.  The current frame, the International Celestial
Reference Frame, will be discussed: its history, the use of the group
delay as the measured quantity, the positional accuracy of 0.3 mas,
and possible future improvements.  On the other hand, for the
determination of the motion of celestial objects, accuracies
approaching 0.01 mas can be obtained by measuring the differential
position between the target object and nearby stationary sources.
This {\it relative} astrometric technique uses phase referencing, and
the current techniques and limitations are discussed, using the results
from four experiments.  Brief comments are included on the
interpretation of the Jupiter gravity deflection experiment of
September 2002.

\end{abstract}
\maketitle
\section{Introduction}

    This paper is based on a talk at the JENAM2003 meeting in
Budapest, Hungary in August 2003, and will cover several topics.  In \S
2, reference systems are defined and a brief historical sketch is given.
In \S 3, the fundamental synthesis formula used for the determination
of radio sources positions, and two basic observed quantities, the
phase and the group delay, are discussed.  The International Celestial
Reference Frame (ICRF) is described in some detail in \S 4.  The
measurement of relative positions to obtain the motion of radio
sources, and calibrator choice concerns are then given in \S 5, and in
\S 6, the description of four experiments highlight current
techniques.  A brief conclusion is given in \S 7, and an appendix
concerning the controversy of the speed of gravity ends the paper.
 
\section{Reference Frames and Systems}

\subsection {Inertial Frames}

    In order to study the position and motion of objects, a reference
frame is needed.  In a normal three-dimensional space (a flat
Minkowski space), three coordinate numbers specify the
location of an object.  These coordinate numbers are defined on a
reference frame which is given by an origin (zero point) and the
direction of two of three orthogonal axes.  In principle, one 
can choose an arbitrary origin and axes directions, but some
reference frames are {\it better} than others.

    If a rotating or accelerating reference frame is used, then the
objects in this frame will have peculiar motions which cannot be
understood by application of dynamical laws to their motions.  For
example, within the earth-centered reference frame (right ascension
and declination), the stars show an annual parallactic motion which
depends on their distance and direction in the sky.  These apparent
motions are, of course, caused by the earth orbital motion; hence our
adopted reference frame was rotating in space.  In fact, the best
determination of the quality of a reference frame is to look for
un-dynamic and strange correlated motions of objects in the frame, and
try to ascertain what frame motion is causing the aberrant behavior of
the stars.  

    Non-rotating and non-accelerating reference frames are called {\it
inertial} frames, and all of the laws of mechanics are
frame-independent as long as an inertial reference frame is used.
But, how can these frame properties be determined without some
knowledge of an {\it absolute space} with no motion and rotation---a
preferred reference frame?  It is generally agreed that such a frame
does not exist according to Mach's principle and Einstein's theory of
general relativity (GR)\footnote{See
http://www.bun.kyoto-u.ac.jp/~suchii/mach.pr.html}; nevertheless, we
can choose a reference frame which is quasi-inertial in that its
origin has negligible acceleration and the reference axes have little
rotation---at least to the accuracy of the measurements.

  For the studies of celestial bodies, the most convenient celestial
reference frame has its origin at the barycenter position of the solar
system.  The motion of the sun around the center of the galaxy, the
gravitational bending from the bulge of our galaxy, and the motion of
the galaxy in space produce both large static corrections (which can
be removed from the source positions) and small differential
corrections over the sky at the 0.01 mas level (Sovers, Fanslow \&
Jacobs \cite{sov98}).  The precise position of the barycenter is in
error by several km because of the uncertainty in the masses of the
major solar system members, predominantly the sun and Jupiter.  The
two axes of the barycenter frame (usually called the pole and
principle plane directions) of the celestial reference frame are
generally defined by the orientation of the earth in space at some
specified time.  Since most observations are made from a slowly moving
crust on the surface of the earth which also orbits the sun and
rotates somewhat non-uniformly around a wobbling axis, the difference
of the earth orientation at the observation time to that at the
fiducial time is a major challenge in obtaining the accurate position
of celestial objects on the adopted quasi-inertial reference frame.

    After defining a suitable reference frame, a method must be
developed by which observations of celestial objects can be located on
this frame.  This is commonly done in two ways: A {\it kinematic
frame} is one in which the positions of a suitable set of celestial
objects are known at any time (some of the distant objects can be
considered fixed in the sky).  Thus, by comparing observations of
target object with these fiducial objects, the target coordinates can
be placed on the adopted reference frame.  In principle, only two
fiducial objects are needed to specify the frame completely, but
observations of both objects and the target will often be impossible;
hence, the number of fiducial objects should be much larger so that
several can be accessible near the time and position of the target
observation.  A {\it dynamic frame} is defined, not by a fiducial set
of objects in the sky, but by an accurate ephemeris of the solar
system bodies and the earth rotation and orientation.  This
information can be obtained, for example, by the motion of the stars
which reflect the earth motion.  In the 1970's and 1980's, the use of
spacecraft tracking, laser-ranging to non-earth surfaces and planetary
radar observations added significant accuracy in the determination of
the orbital, rotational and orientation motions of the earth.  Of
course, the realization of a reference frame can consist of a
combination of both kinematic and dynamic information.

\subsection {Previous Reference Frames}

    Newcomb's studies of the relative motion of stars in the $19^{th}$
century produced accurate values of the precessional motion of the
earth's pole, the length of a day and year, by measuring the transit
time of about 1000 bright cataloged stars.  With these constants, the
FK3 catalog (Fundamental Katalog 3; Kopff \cite{kop38}) of the star
positions and the assumed precession and nutation constants defined a
celestial reference system which could be used to determine the
coordinates of any star or solar system object to an accuracy of
$2''$.  The extension of catalogs to more than 10000 bright stars and
more accurate proper motion determinations led to the compilation of
the FK4 catalog (Fricke \cite{fri63}) along with better precessional
constants\footnote{
http://www.to.astro.it/astrometry/Astrometry/DIRA2/DIRA2\_doc/FK/FK4.HTML}.
The accuracy of this improved reference frame increase to about
$0.5''$, with some degradation in the southern sky.  Both frames are
dynamical because they relied on the modeling of the earth's orbital
and spin motion using the residual motion of many stars.

    The FK4 system was not keeping up with the accuracy of the
observations, and created pressure for continuing revision of the
fundamental constants of the system.  Thus, the FK5 system (Fricke
\etal~\cite{fri88}) was developed around 1980.  The procedures which
improved the FK5 frame were better measurements of the proper motion
of thousands of stars, and the specific assumption that distant
galaxies were fixed in the sky.  This added a kinematical frame
component to FK5 reference frame definition and improved the accuracy
to about $0.2''$ by 1990.  Other astronomical observations---the
motion of near-earth asteroids and occultations of stars by
planets---also increased the accuracy of this frame.  But, the seeds
were planted for a kinematical approach to future reference frames.

    Beginning around 1980, the Jet Propulsion Laboratory (JPL)
determined a celestial reference using many types of observations,
mostly of the dynamic type.  The data were gathered from: optical
observations of the planets, Viking spacecraft range observations,
radar observations of Mercury, Venus and Mars, Lunar-laser ranging,
asteroid perturbations, and precision lunar
modeling.  This DE200 ephemeris was tied to the FK5 system by proper
overlap of planetary objects (Standish \cite{sta90}).

\subsection {Current Reference Frame}

    The technique of Very Long Baseline Interferometry (VLBI)
demonstrated in the 1970's that many radio sources have significant
emission within a component (radio core) less than 1 mas.  These
sources are identified with quasars (galaxies with intense point-like
stellar nucleii) at distance on the order of 1 Gpc.  Thus, it is an
excellent assumption that these sources are virtually fixed in the
sky, and they could define an inertial reference frame to much higher
accuracy than the optical-based FK5 frame or the planetary
ephemerides.

Over the next 20 years, improvements in the stability and observing
procedures of VLBI, with more precise modeling of the earth rotation,
nutation, orientation and planetary motions, led to quasar position
accuracy approaching 1 mas over the sky.  In the early 1990's the
astronomical community formed working groups in order to define an
International Celestial Reference System (ICRS) (Ferraz-Mello
\etal~\cite{fer96}; Feissel \& Mignard \cite{fei98}) which could best
utilize the accurate astrometric theory and results.  The ICRS was a
set of rules and conventions, with the modeling required, to define at
any time the orientation of the three coordinate axes (only two are
independent), located at the barycenter of the solar system.  The axis
directions were fixed relative to a suitable number of distant
extragalactic sources.  For reasons of continuity with the FK5 system,
the ICRS-defined pole direction at epoch J2000.0 was set to the FK5
pole at that epoch, and the origin of right ascension was defined by a
small radio component in the source 3C273 with an accurate position
measurement determined by a lunar occultation (Hazard
\etal~\cite{haz71}).  The catalog of positions of the fiducial objects
needed to realize the ICRS is called the International Celestial
Reference Frame (ICRF) (Ma \etal~\cite{ma98}).

Since many telescopes around the globe participate in VLBI
observations, a complementary terrestrial reference frame was needed.
Thus, the International Terrestrial Reference System (ITRS) was
formulated to describe the rules for determining specific locations on
or near the earth.  In order to realize the ITRS, the international
Terrestrial Reference Frame (ITRF) consists of a catalog of about 50
fiducial locations of the earth.  Global Position Service (GPS)
observations over the last 10 years have also provided high accuracy
in the determination of the ITRF (see
http://www.iers.org/iers/products/itrf).

\section {Position Determination from Radio Array Observations}

    This paper will describe two aspects of astrometry: all-sky {\it
absolute} astrometric techniques to determine the accurate positions
used in defining the ICRF, and {\it relative} astrometric techniques
used to determine the motion of individual objects.  However, a brief
introduction to radio interferometry is needed in order to understand
the calibration methods and the different observing strategies used
for absolute and relative astrometry.

\subsection {The Phase Response of an Array}

    A radio source signal, intercepted by two telescopes pointed in
the same direction of sky, is essentially identical except for the
different travel time (delay) of the radio wave from the quasar to
each telescope.  The correlation (multiplying and averaging) of the
two signals produces a response called the spatial coherence function.
Its amplitude is related to the strength of the source and its angular
size; its phase is related to the delay of the signal between the
telescope, with a minor contribution from the source structure.  For
an array of many telescopes, each telescope-pair (baseline) gives an
independent measure of the spatial coherence function.

    Using a~priori calibrations of the telescope amplification
properties, the coherence amplitude can be converted into true energy
units (flux density), and is denoted as the visibility amplitude.  The
coherence phase is a rapidly changing function of time because the
earth rotation changes the delay between each baseline.
However, an accurate model of the parameters describing the
observations (the location of the telescopes on the earth
surface, the orientation and rotation of the earth in space, the
position of the radio source, propagation delays in the troposphere
and ionosphere, etc) can be calculated at any time for any baseline.
When this model coherence phase is subtracted from the observed
spatial coherence phase, a slowly changing residual phase, the
observed visibility phase, is obtained.

    The basic components of the observed visibility phase,
$\phi^a_{l,m}(t)$ observed at frequency $\nu$ for source $a$ at time
$t$, between telescopes $l$ and $m$ are

\begin{eqnarray}
   \phi^a_{l,m}(t) = \psi^a_{l,m}(t) + \frac{\nu}{c}\Bigl[
        \big({\bf R}_l-{\bf R}_m\big){\bf \cdot} {\triangle{\bf K}}^a(t) \nonumber \\
      + \big({\triangle}{\bf R}_l(t)-{\triangle}{\bf R}_m(t)\big)\cdot{\bf K}^a +
        \big(C_l(t)-C_m(t)\big) + \big(A^a_l(t)-A^a_m(t)\big) \Bigr] \nonumber \\
   \label{eq1}
      + \frac{1}{\nu c}\Bigl[I^a_l(t)-I^a_m(t)\Bigr]
           \hbox {;~~~~~~~~~~~~with~~}-0.5 < \phi^a_{l,m}(t)  <+0.5  
\end{eqnarray}
\noindent
where the structure phase of the $ath$ source is given by
$\psi^a(l,m)$, and is zero for a point source.  ${\bf R}_l$ is the
model location of telescope $l$, and $\triangle{\bf R}_l(t)$ is the
unknown location offset; ${\bf K}^a$ is the model position of the
$a$th source, and $\triangle{\bf K}^a(t)$ is the unknown position
offset, $C_l(t)$ is the residual instrumental and clock delay error
for telescope $l$, and $A^a_l(t)$ is the residual tropospheric
propagation delay in the direction to the $ath$ source for telescope
$l$.  The residual ionospheric refraction for telescope $l$ is
$I^a_l(t)$.  Since this delay is produced by plasma refraction, the
phase varies as $\nu^{-1}$.  Those quantities in bold face are vectors
on the ground or directions in the plane of the sky.  However, {\it
the measured visibility phase is only defined between $-0.5$ and
$+0.5$ turn.}

   The first line in Eq.~(\ref{eq1}) gives the terms which are
associated with the source properties: its structure and position.
When combined with the visibility amplitude, the ensemble of
visibility amplitudes and phases for all baselines and times is the
two-dimensional Fourier-pair of the source brightness distribution in
the sky (Thompson, Moran \& Swenson \cite{tho94}).  Hence, the
Fourier transform of the visibility amplitude and phase will produce
an image from which positional information can be obtained.

    The second line in Eq.~(\ref{eq1}) shows the major error terms
which have significantly different spatial and temporal properties.
For example, $\triangle {\bf R_l}$ represents the slowly changing
location of the telescope (with respect to the a~priori model) which
are caused by continental drift, or earth orientation, rotation and
nutation uncertainties.  This error term (by virtue of the dot product
with the source position) has a period of 24 hours.  The $C$-term
represents all temporal delay errors associated with a telescope,
including that from the independent maser clocks, with a typical one
hour time variation time scale .  The propagation delay through the
troposphere $A$ (or the ionosphere $I$ at low observing frequencies)
is the most intractable of the error sources.  Even with detailed
ground meteorological measurements and tropospheric/ionospheric
path-length monitoring from GPS satellites, the a~priori delay model
can be significantly different than the actual path delay above each
telescope.  See Treuhaft and Lanyi (\cite{tre87}), Niell
(\cite{nie96}) and MacMillan \& Ma (\cite{mac97}) for more information
on modeling the troposphere.

    In order to obtain {\it absolute} positions, all of the error
contributions must be determined and parameterized as accurately as
possible, and the necessary observing strategies are described in \S
4.  On the other hand {\it relative} positions are generally obtained
by alternating observations of the target with a nearby fixed radio
source (calibrator) with known properties, and assuming that the
errors affecting the observations of the calibrator are nearly the
same as that for the target source.  This technique is described
in \S 5.

\subsection {The Group Delay}

    For small arrays ($<100$ km), the experimental a~priori model is
sufficiently accurate so that cycle ambiguities of the phase are not a
problem, and Eq.(~\ref{eq1}) can be used directly to obtain accurate
positions (Wade \cite{wad70}).  However, for most VLBI experiments
with baselines well in excess of 1000 km, even the most accurate a
priori model now available produce residual delays at telescope
baselines of 5000 km which are $>100$ psec, equivalent to a 4 cm
path-length, more than one cycle of phase at $\nu=8~$GHz.  At
frequencies above 2 GHz, the dominant error is caused by the
troposphere; at lower frequencies the ionospheric delay becomes
dominant, and delay changes over short times scales and in small
patches of sky can exceed 5 cm.  Thus, the measured visibility phases
have cycle ambiguities and are difficult to use directly for
astrometric analysis.

The phase ambiguity problem can be overcome by the measurement of the
derivative of the visibility phase with frequency, which is called the
group delay, $G=d\phi/d\nu$.  It can be determined by observing at
many frequencies simultaneously, or nearly simultaneously, to derive
the phase slope.  As long as some selected observing frequencies
are not too widely separated, the group delay is not affected by phase
cycle ambiguities.  The use of the group delay for astrometric work
and the need for special calibrations and frequency sampling were
first described in detail by Rogers (\cite{rog70}), and the technique
is often called bandwidth-synthesis.  The derivative of
Eq.~(\ref{eq1}) with respect to $\nu$ is trivial and simply removes
the $\nu$ in front of the large bracket.  Two additional terms are
produced by source structure (Sovers \etal~\cite{sov02}), which causes
errors at the level of 5 psec, and the ionospheric refraction which
can be removed by observing with a wide frequency range.  Hence, the
solution form using the phase or the group delay is
identical\footnote{The phase derivative with time called the {\it
rate}, which also does not contain cycle ambiguities with short time
sampling, can also be used to obtain astrometric solutions.  The
functional form of Eq.~\ref{eq1} upon differentiation with time is
more complicated.  The phase rate is less accurate than the group
delay because of the short-term contamination by the tropospheric
phase changes, although at low elevations the rate term is a important
contribution to the data.  The rate is also useful for narrow-band
line emission which have poorly defined group delays.}.  However, the
group delay can only be adequately measured for relatively strong
sources for which an accurate visibility phase can be determined in
most of the simultaneously-observed channels.  Astrometry using weak
sources is not possible (see \S 5 for more details).

\section {The ICRF and Absolute Astrometry}
\subsection{Description of the Observations and Reductions}
    The majority of the VLBI observations used for the ICRF comes from
observations by the NASA Crustal Dynamics Project (CDP), the United
States Naval Observatory (USNO), the Jet Propulsion Laboratory (JPL),
and other groups.  Although astrometric observations began as early as
1971 at JPL, the wide-band, dual-frequency observations which are the
basis of the ICRF solutions started in 1979.  This set of data will be
collectively designated in this paper as CDP observations.  The
observations were made simultaneously at 2.3 GHz and 8.4 GHz with at
least four frequency channels (each with 16 subchannels or 16 delay
lags) at both frequencies.  The spanned bandwidth at 2.3 GHz was about
0.1 GHz, and at 8.4 GHz about 0.4 GHz, from which accurate group
delays were determined.  By a suitable combination of the
two-frequency data, the ionospheric refraction was
determined and removed, to produce 8.4 GHz data which is then
ionospheric-free (Sovers, Fanslow \& Jacobs \cite{sov98}).

The observing strategy was carefully planned in order to obtain robust
solutions for all of the unknown parameters (Ma \etal~\cite{ma98}).
Each observing period (denoted as a session) lasted 24-hours in order
to fully sample the $\triangle{\bf R}\cdot{\bf K}$ terms in
Eq.~(\ref{eq1}).  By about 1990 each session had evolved into a
schedule which consisted of about 150 observations in which about 50
good quality sources were observed for one to ten minutes
with no set number of scans per source.  Sources were scheduled over
the entire sky in a relatively short time in order to determine the
variable tropospheric refraction which produces the largest source of
error.  Determination of each telescope clock delay and mean
atmosphere zenith-path delay and gradient were determined at hourly
intervals.

    Many different arrays and telescopes have been used over the
years, but the observational plan has remained essentially unchanged.
The group delay and rates are analyzed by several available software
packages (Modest--Sovers, Fanslow \& Jacobs \cite{sov98},
Calc/Solve--Ryan \etal~\cite{rya93}) in order to obtain all of the
parameters associated with the experiment.  Some parameters are
global, unchanged over the entire period covered by the observations
(source positions, some telescope properties), some parameters are
slowly variable but considered as constant over each 24-hour session
(precession/nutation, UT1 offset and rate, telescope locations), and
some parameters are extremely variable with parameter estimates made
every hour (zenith path delays and gradients, telescope clock drifts).
The typical residual delay error for each 24-hour session after
obtaining the best solution is about 25 psec, and is dominated by
residual tropospheric delay residuals.  This corresponds to a phase
error at 8 GHz frequency of $80^\circ$.

\begin{figure}
  \includegraphics[width=6.2cm]{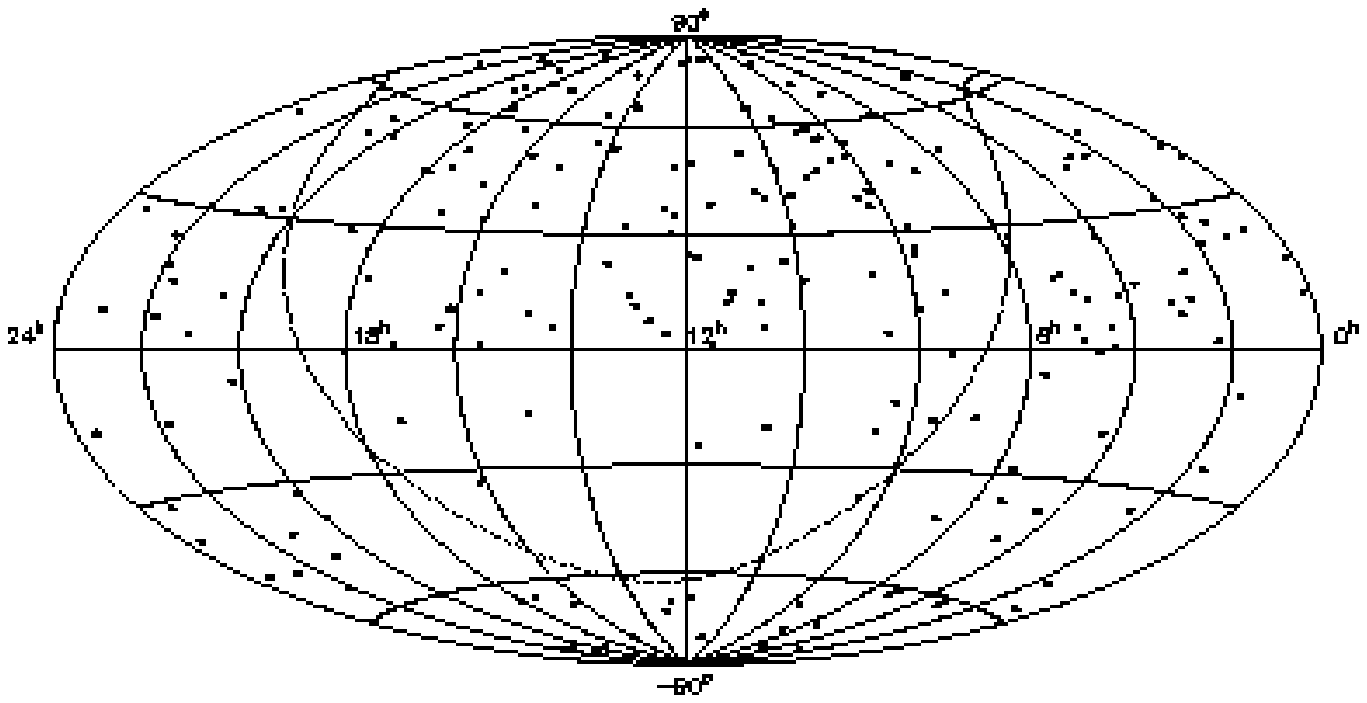}
  \includegraphics[width=6cm]{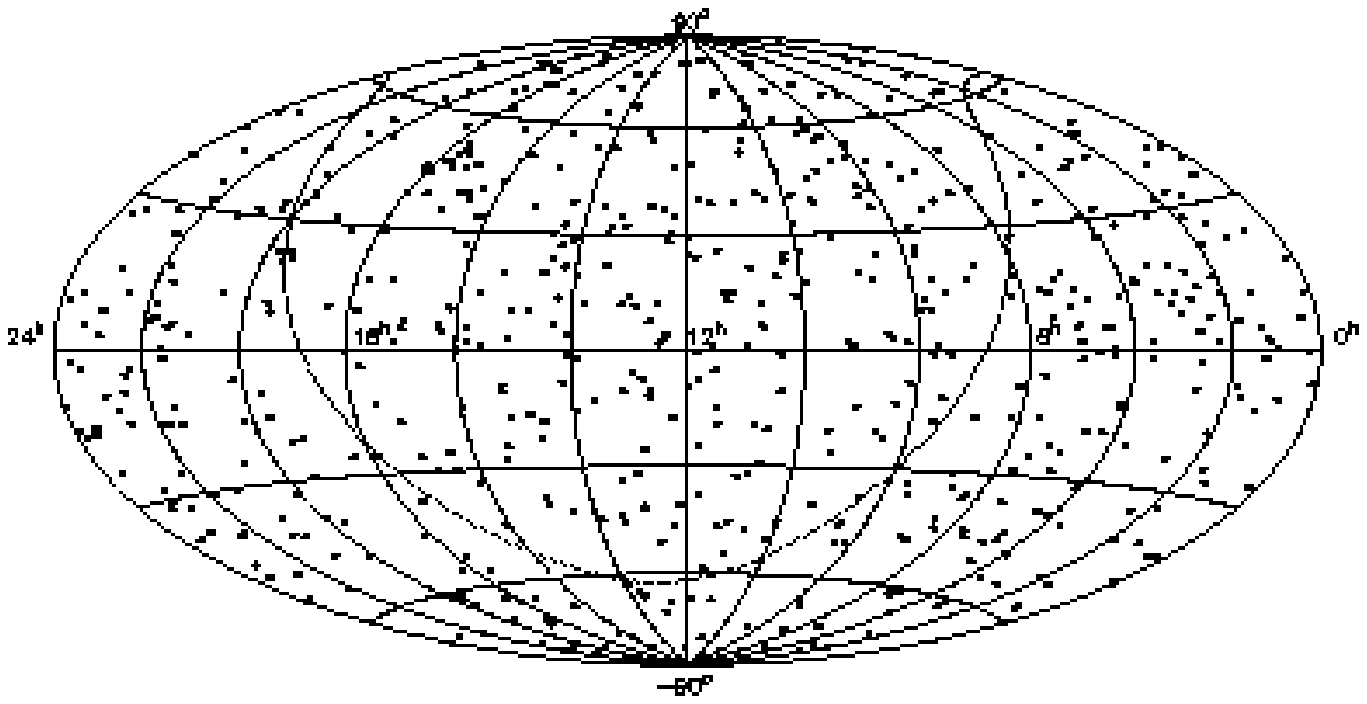} \caption{{\bf (left)} The sky
  distribution of the 212 defining sources in the ICRF (from Ma~\etal~1998
Fig.~10); {\bf (right)} the
  sky distribution for all sources in the ICRF (from Ma~\etal~1989, Fig.~13).}
\end{figure}

\begin{figure}
  \includegraphics[width=8cm]{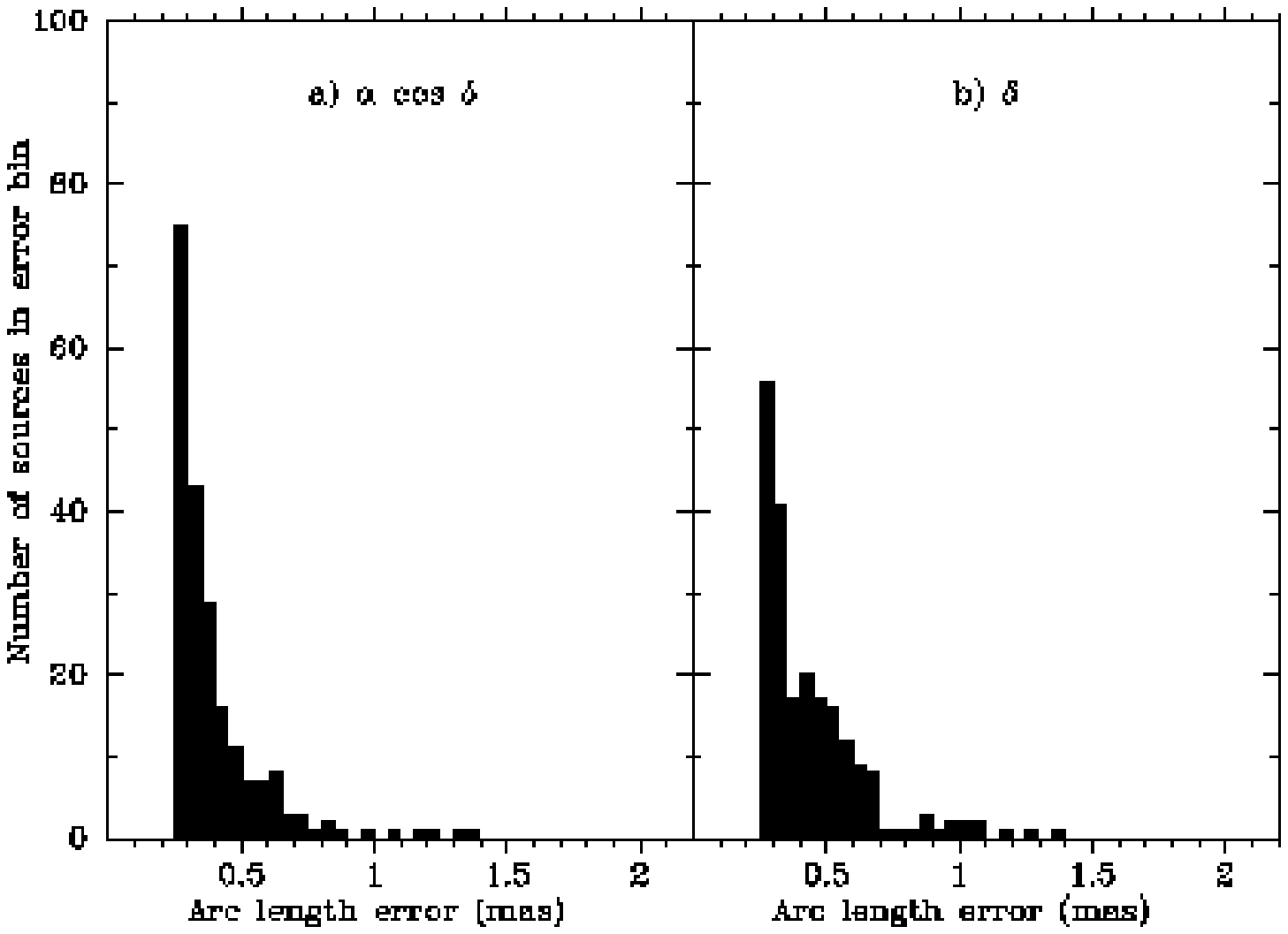}
  \includegraphics[width=4cm]{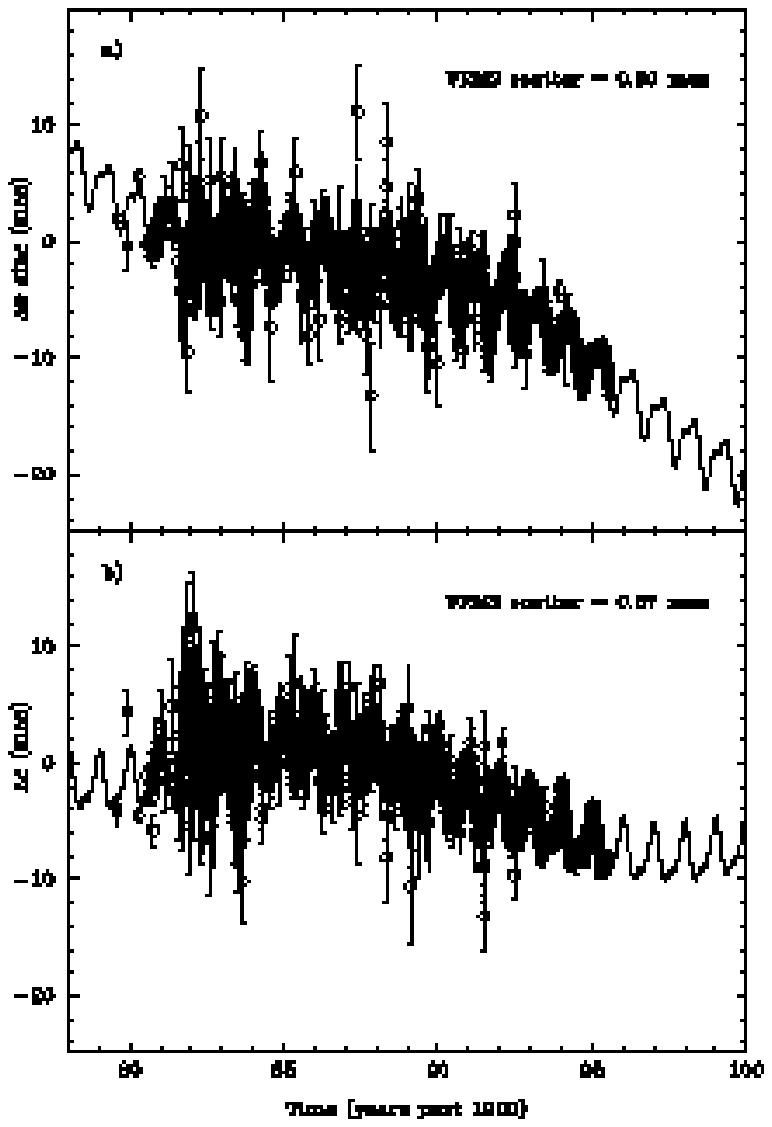} \caption{{\bf (left)} The
  histogram of source position errors for the defining sources in
  $\alpha~$cos$\delta$ and $\delta$ (from Ma~\etal~1998, Fig.~6a and
  6b). {\bf (right)} The measured nutation terms, with respect to the
  IAU 1980 model, observed for each session.  The plotted curve is
  fit to the data using known nutation periods with 18 yr, 9 yr, 1 yr,
  0.5 yr, and 14 days period (from Ma~\etal~1998, Fig.~1a and 1b).}
\end{figure}

   In the fall of 1995, a combined solution of all previous CDP
observations determined thousands of parameters.  The global
parameters of the source positions and some telescope properties were
held fixed over the whole set of data.  The details concerning this
complex least-squares fit to determine the most consistent set of
source positions are given elsewhere (Ma \etal~\cite{ma98})\footnote
{http://hpiers.obspm.fr/webiers/results/icrf/icrfrsc.html}.

    The resulting catalog contains three sets of sources.  The `best'
212 sources have frequent observations, relatively small position
errors, and are dominated by a small-diameter radio component.  They
were used to define the orientation of the axes of the ICRF.  In
addition, there are about 300 sources with a smaller number of
observations, less accurate positions, or obviously unstable positions
(due to large source structure changes).  About 100 sources with
bright optical counterparts were included for potential frame-tie
connections.  The sky distribution of the defining 212 defining
sources in the ICRF is shown in Fig.~1 (left).  The distribution is
relatively evenly spread north of the equator.  The distribution for
the entire set of ICRF sources is given in Fig.~1 (right).
Observations are continuing and an additional 59 good quality sources
have been added to the ICRF-extension1 catalog\footnote{
http://hpiers.obspm.fr/webiers/results/icrf/icrfext1new.html}.

    The distribution of the root-mean-square (RMS) radio source
position error is given in Fig.~2 (left).  In order to obtain
realistic, but conservative error estimates, the internal errors,
generated from the scatter among the sessions, were multiplied by 1.5
and then added in quadrature to a 0.25 mas error.  These errors should
then include structure effects over time and frequency.  The accuracy
of the ICRF realization of the ICRS axes is about 0.02 mas.  An
indication of the accuracy of the CDP observations can be seen in
Fig.~2 (right) which shows the derived correction to the pole
orientation (nutation) obtained for each 24-hour CDP session.  It was
fit to a realistic model known from geophysical and gravitational
considerations with an RMS of the fit of about 0.3 mas.  This is
another indication of the accuracy of the frame realization for a
24-hour long session.

\subsection{Frame Ties}

   Since the ICRF frame is accurate at the level of a few 0.02 mas
and is stable over time scales of decades, it is advantageous to link
other frames to it.  The link to the FK5 frame was discussed in
connection with continuity of the direction of the pole and origin of
right ascension, with the ICRF adopting the FK5 pole at a specific
epoch of time.

    The Hipparcos catalog (Perryman \etal~\cite{per97}) contains
observations between 1989.85 and 1993.21 of over 100,000 stars
brighter than 12-mag.  These observations were placed on the Hipparcos
Reference Frame (H{\o}g, E.~\etal~\cite{hog97}) to an accuracy of 1.0
mas.  The link between the Hipparcos optical frame and the ICRF was
accomplished in several ways.  First, the few bright stars with
significant radio emission were observed with the VLBI, the VLA and
with Merlin in order to determine the radio positions with respect to
the ICRF (Lestrade \etal~\cite{les95}).  Accurate parallaxes and
proper motions were also obtained for these radio emitting stars
(Boboltz \etal~\cite{bob03}).  Secondly, comparison of the Hipparcos
optical with HST observations linked these stars to extragalactic
objects, and then to nearby ICRF objects which could be detected by
HST.  Next, additional optical observations using ground based
telescopes also helped link Hipparcos positions to nearby galaxies.
Finally, the earth orientation parameters were obtained from the
Hipparcos observations and then compared with that obtained from that
from nearly concurrent VLBI observations (Lindegren \& Kovalevsky
\cite{lin95}).  With these comparisons, the Hipparcos optical catalog
is now consistent with the ICRF to about 0.6 mas in position and 0.25
mas/yr for axis rotation.  However, the position of an individual star
and the reference frame tie decreases in accuracy with time because of
the proper motion uncertainties of the stars of 1 mas per year.

   The link of the ICRF to the planetary dynamic ephemeris was
accomplished using several techniques.  First, the time of arrival of
a pulsar signal is sensitive to the orbital motion of the earth.  By
comparing the pulsar position derived from timing analysis with that
obtained with VLBI observations, improvements in the dynamics of the
earth motion could be obtained (Dewey \etal~\cite{dew96}).  Secondly,
occultations of radio sources with solar system objects also tie the
ICRF to that of the planetary ephemeris.  Finally, radio
interferometry of spacecraft with respect to nearby ICRF source also
improved the frame tie between the two systems; for example, the
observations the Magellan and Pioneer spacecraft with respect to
nearby quasars (Folkner~\etal~\cite{fol93}).  Comparison of the lunar
ranging with VLBI earth orientation results were also important in
tying the two frames together (Folkner~\etal~\cite{fol94}).  The
resulting JPL DE405 ephemeris is connected to the ICRF to about 1.0
mas in position and about 0.05 mas/yr in rotation (Standish
\cite{sta98}), and the tie has recently been improved by a factor of
two or three (Border 2003, private communication)

\subsection{The Future of the ICRF}

   The ICRF will be the best realization of a quasi-inertial system at
least until 2015 when the Space Interferometry Mission (SIM) has been
in operation for several years.  Hence, improvements in the ICRS and
the accuracy of the ICRF positions are needed, and may be obtained in
several ways.  First, many more observations of the ICRF sources have
been made since the original solution was made in 1997, and improved
positions are designated as the ICRF-ext1 catalog.  A further update,
ICRF-ext2, is expected (Fey~\etal~2004, in preparation).  Secondly,
VLBA observations, in its search for suitable calibrators needed for
phase referencing (Beasley \etal~\cite{bea02}; Fomalont
\etal~\cite{fom03}), have produced a catalog of over 2000 radio
sources, some of which are candidates for additions to the ICRF
catalog.  Also, the number of experiments for sources in the southern
hemisphere (Roopesh \etal~\cite{roo03}) has enlarged over the last ten
years.  A new defining list of ICRF sources may be generated in the
next five years, with an expected improvement in the grid accuracy of
about a factor of two.  However, discontinuities between the old and
new frame definitions must be avoided.

\begin{figure}
  \includegraphics[width=6.5cm]{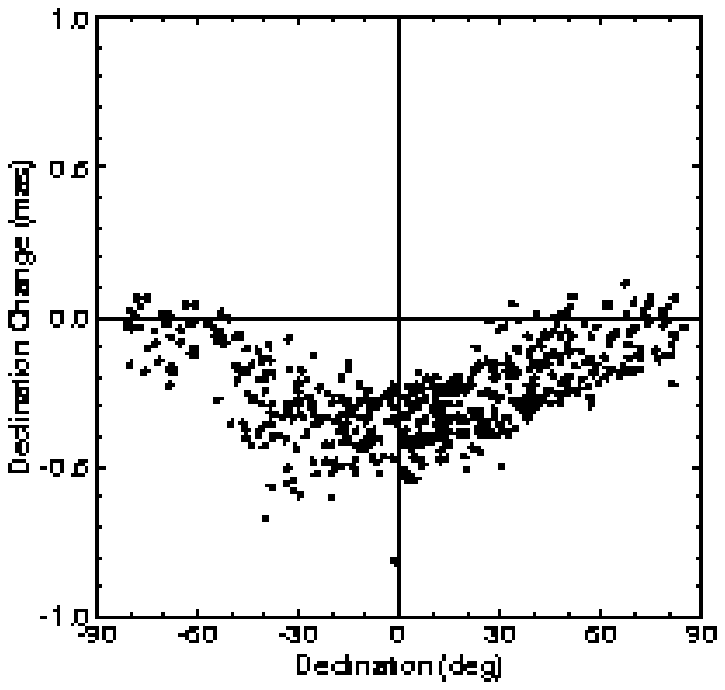} \qquad
  \includegraphics[width=4.5cm]{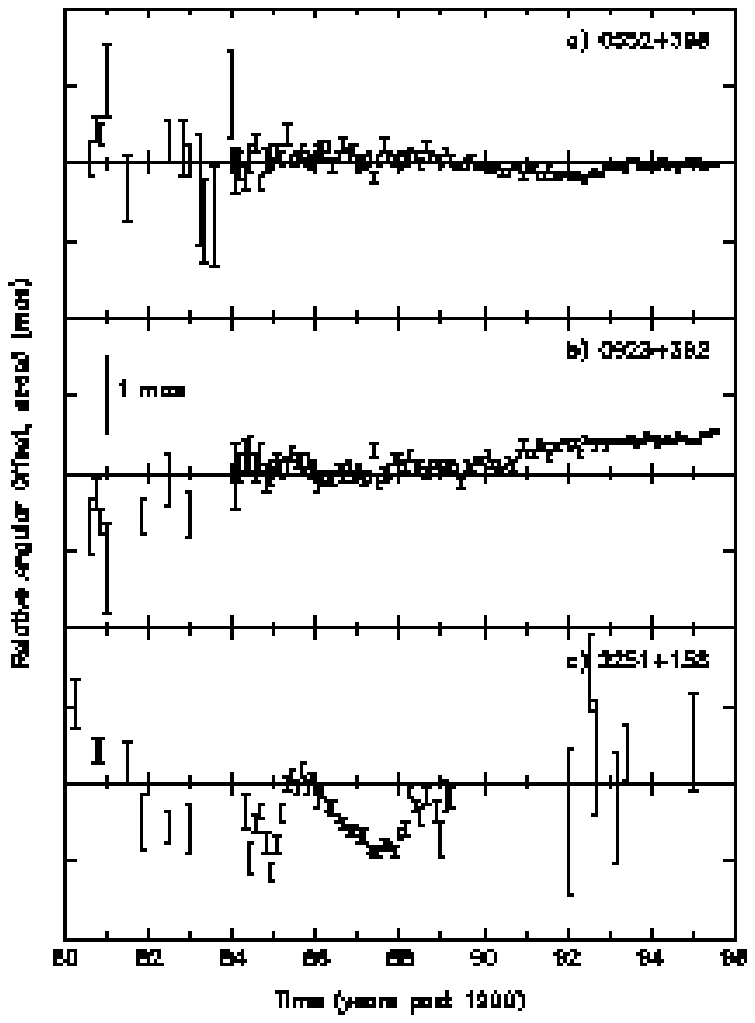} \caption{{\bf (left)}
  The difference in the declination of the ICRF sources, as a function
  of declination, using global solutions with and without a
  tropospheric gradient in the modeling (from Ma~\etal~1998,
  Fig.~4). {\bf (right)} The change of position of three ICRF sources
  as a function of time caused by moving emission substructure near
  the radio core (from Ma~\etal~1998, Fig.~2).  The abscissa scale
  covers eight years and 1 mas is indicated in the middle plot.}
\end{figure}

    A major source of error in the parameters associated with the ICRF
is caused by the variable tropospheric delay; it is the major
contributor to the current limit of 0.25 mas accuracy for the good
quality sources.  Although the troposphere zenith path-delay and
gradient are modeled every hour during most 24-hour sessions, the
structure and kinematics of the troposphere are extremely complicated,
and the typical RMS delay scatter, after obtaining the best solution,
is about 25 psec at a 5000-km baseline, which corresponds to a
positional error of about 0.2 mas at 8 GHz.  More sophisticated
models, better monitoring of the tropospheric delay using GPS
satellites, and ground measurements of water vapor emission in the
line of sight to a source could decrease this uncertainty to one-half
of the present level, but progress has been slow.  An example of the
gain in precision by an improvement in the tropospheric modeling is
shown in Fig.~3 (left).  The plot shows the difference in the derived
declination of sources between two reductions of all of the 24-hour
sessions in 1995: one reduction used an azimuthally-symmetric
tropospheric delay model over each telescope and the other included a
tropospheric gradient term (which is now routinely used).  The
differences as a function of source declination are up to 0.4 mas and
show that systematic errors would have been present in the ICRF frame if
the simpler tropospheric model had been used.  Any remaining
systematic error could be as large as 0.1 mas.

    Another source of error is associated with the evolution of
quasars---their changing structure.  While these distant objects are
fixed in the sky, the radio emission does not emanate from the
galactic nucleus, but from the inner jet region which may be located
0.1 mas from the core at 8 GHz, and could be somewhat variable in
position.  Another source of error is caused by radio-emitting clouds
of magnetic plasma which propagate down the jet and are often much
brighter than the radio core nearer the galaxy nucleus.  Fig.~3
(right) shows the apparent position versus time for several quasars
which have undergone such evolution.  The typical time-scale of
changes are several months to years, and apparent position movements
of 0.1 mas are common, and occasional changes as large as 1 mas can
occur.  With our knowledge of the nature of quasar radio emission, and
with frequent monitoring of the radio source emission structure, it
should be possible to model these evolutionary changes to remove the
effects of these apparent positional changes (Fey \& Charlot
\cite{fey97}; Sovers \etal~\cite{sov02})).

    Finally, the ICRF positions are based on radio source properties
at 8.4 GHz, but observations and studies are now underway to define a
complementary ICRF at a frequency of 32 GHz
(Lanyi~\etal~\cite{lan03}).  At this higher frequency, the ionospheric
refraction is smaller and the radio source structures are more
compact.  However, radio variability and structure changes will still
be a problem.  NASA-JPL plans to use 32 GHz for spacecraft telemetry,
beginning in 2005.  With an ICRF defined at this frequency, the tie
between the planetary ephemerides and the quasar reference reference
could be strengthened.  However, such a tie is now perturbed by the
limited modeling of the asteroids at the level of 0.2 mas per year
which may be the dominant error (Standish \& Fienga \cite{sta02}).

    The advances in absolute astrometric measurements over the next 20
years (including space interferometry) may reach accuracies at the
level of 0.01 mas, when a true space-time reference frame must be
considered.  This means that the astrometric system must become
four-dimensionally based rather than the current system which is three
dimensional, but with correction factors associated with the variable
gravitational field of the solar system.  Also, there is some
discussion that the origin of the ICRF should be moved to the
barycenter of the earth-moon system, rather than that of the solar
system, because of the lower gravitation potential and diminished
gravitational space distortion near the earth compared with that near
the sun (Soffel \etal~\cite{sof03}).

\section {Relative Astrometry}

    Many astrophysical phenomena are associated with the accurate
space motion of celestial objects, as well as their accurate position.
For a source in the Milky way, the parallax is the only direct method
of determining the distance, and its proper motion is often related to
the evolution and formation of the object.  From the orbital motion of
a source, many properties of a binary or multiple system, including
the detection of planet-like objects, can be made.  For solar system
objects the accuracy of radio interferometry is similar to that of
lunar-laser ranging and range/Doppler measurements, and is
complementary since interferometry determines positions on the
celestial sphere and the other techniques determine the distance.
Finally, the effect of the gravitational field in the solar system on
the propagation of radio waves can be measured and compared with that
predicted by GR or other theories of gravity.
 
\subsection {Target-Calibrator Phase Referencing Methods}

    The determination of the motion of a radio source requires less
complicated observations and reductions than that associated with
all-sky ICRF observations.  The reason is that the motion of a radio
source can be obtained by measuring its position with respect to any
detectable radio source, called a calibrator, that is {\it fixed} in
the sky.  If the calibrator-target separation in the sky is small and
observations are switched rapidly between the calibrator and target,
then the major error terms, shown in Eq.~(\ref{eq1}), are similar for
the two sources and nearly cancel when their observed visibility
phases are differed.  This observation scheme is called {\it phase
referencing} (Beasley \& Conway \cite{bea95}).

From Eq.~(\ref{eq1}) the phase difference, $\triangle\phi^{s-c}_{l,m}$
at any time (t) between the target source $s$ and the calibrator $c$,
for the baseline between telescopes $l$ and $m$ is
\begin{eqnarray}
 \label{eq2} 
   \triangle\phi^{s-c}_{l,m}(t) = \psi^s_{l,m}-\psi^c_{l,m} +
   \frac{\nu}{c}\Bigl[ \big({\bf R_l}-{\bf R_m}\big){\bf \cdot}
   \big({\triangle{\bf K}}^s(t)- {\triangle{\bf K}}^c(t)\big)
   \nonumber \\ 
   + \big({\triangle}{\bf R}_l(t)-{\triangle}{\bf
   B_m(t)}\big) \cdot{\bf\delta K}^{s-c} +
   \partial^2\big(C_l(t)-C_m(t)\big)/\partial t^2 \nonumber \\
   + \delta A^{s-c}(t)\big] + \frac{1}{\nu c}\delta I^{s-c}_{l,m}(t)
\end{eqnarray}
The first line shows the phases which dependent on the target and
calibrator source structures and positions.  The next two lines
contains the differential delay errors.  The small separation of the
two sources, $\delta{\bf K^{s-c}} \equiv {\bf K^s-K^c}$, significantly
decreases the effect of the telescope-location errors (in the most
general sense), for example, by a factor of 50 for a one degree
target-calibrator separation.  Similarly, $\delta A^{s-c}\equiv
A^s-A^c$ is the difference of the tropospheric delay error in the
direction of the two sources, and $\delta I^{s-c}\equiv I^s-I^c$ is
the ionospheric delay.  The quasi-random short-term delays from small
clouds will not cancel particularly well, even for close calibrators,
but larger angular-scale deviations from the a~priori model will
cancel.  This term will be discussed in more detail using
multi-calibrator observations.  Finally, the purely temporal (mostly
clock) delay variations would cancel precisely if the calibrator and
target were observed simultaneously\footnote{If the calibrator and
target are sufficiently close, they can be observed simultaneously. An
existing array, VLBI Exploration of Radio Astronomy (VERA), has been
designed in order to observe two sources, separated by no more than
$2.5^\circ$, simultaneously (Honma \etal~\cite{hon03}). }.  Otherwise,
time-interpolation between calibrator observations is needed, and
small second order clock errors will remain in the difference.  The
switching time between observations of the calibrator and target
depend on temporal characteristics of the $C$ and $A$ terms in
Eq.~(\ref{eq2}).  The longest time for which interpolation between two
observations will produce an accurate phase is called the {\it
coherence} time, and varies from 30 sec at 23 GHz to 5 min at 1.4 GHz.
During periods of inclement whether or strong ionospheric activity,
the coherence time can decrease to 10 sec, making phase referencing
impossible unless the calibrator and target are observed
simultaneously.

\subsection {Phase Versus Group Delay}

    The differential delay error between the calibrator and target are
often less than 5 psec.  Even at a high frequency of 23 GHz, the
corresponding phase difference in $\triangle\phi^{s-c}_{l,m}(t)$ is
less than one cycle and ambiguities of the phase are unlikely to
occur.  (Changes more than one cycle can be used as long as the phase
between subsequent calibrator observations are connected properly.)
The group delays can still be obtained by observing at several
frequencies; however, this is not recommended, unless necessary, for
several reasons.  First, the precision of the visibility phase is
greater than that of the group delay by the factor $\nu/\Delta\nu$
where $\nu$ is the observing frequency and $\Delta\nu$ is the spanned
frequency range used to determine the group delay.  For a reasonably
strong sources, the typical delay error using the measured phase is
$<1$ psec, while that for the group delay is about 10 psec.  Since the
all-sky observations, described for the CDP observations have typical
residuals of 25 psec, which are dominated by the troposphere errors,
the use of the much more accurate phase data (if it were possible to
sort out cycle ambiguities) would not provide more accurate solutions
than use of the group delay.  On the other hand, the delay errors from
a phase-referencing experiment are often $<5$ psec, which is less than
the inherent group delay accuracy, but not the phase accuracy.

   Another reason for using the phase, rather than the group delay, is
to preserve the imaging capability afforded by the visibility phase.
If the error terms in the second line of Eq.~(\ref{eq2}) can be
determined (or assumed to be negligible), then Fourier imaging of the
residual phase will produce an image of the source with its position
relative to that of the calibrator.  Even if the target source is
extremely weak and cannot be detected during a single observation,
such Fourier imaging of the entire target data set will produce an
image from which accurate astrometric information can be obtained.
For these weak sources, the group delay cannot be measured.

\subsection {Calibrator Properties}

    The two most important properties of a calibrator source are its
proximity in the sky to the target, and the strength of a compact
radio component which must be detectable within a coherence time.  In
order for the calibrator phase to be accurately determined in a few
minutes, the radio core must contain no less than ~20 mJy (using the
VLBA at 8 GHz at 64 MHz bandwidth which gives an RMS noise of 4
mJy).  The use of a much stronger calibrator source which is more
distant from the target than a fainter, but otherwise suitable
calibrator, is not recommended.  The decrease in the measurement phase
error from a stronger source is more than balanced by the increase in
differential errors between the calibrator and target.

    Many calibrators contain very extended emission and even most
compact radio cores are not true point sources.  The phase effect of
this structure is contained in the $\psi^c$ term in Eq.~(\ref{eq2}).
Using the self-calibration techniques for which images of strong
sources can be made (Cornwell \& Fomalont \cite{cor99}), the structure
phase can be obtained and removed.  Thus, imaging of all calibrators
should be done routinely.

    A more serious problem is the variable position offset of the
calibrator $\triangle{\bf K^c}(t)$ between observing sessions, since
this change is translated directly into an apparent position change of
the target.  This problem is associated with extended sources where
the maximum radio brightness or centroid brightness moves along the
jet and with respect to the (fixed) galaxy nucleus (see Fig.~3 (right)
for some examples).  Even for compact sources, it is possible that its
position shifts with time since the radio emission probably emanates
from the inner part of the radio jet which may not be a stationary
position with respect to the galaxy nucleus.  When trying to reach
astrometric levels $<0.1$ mas in a target over a period of time longer
than a few months, it is crucial to image the calibrator in order to
determine the probable angular change from the calibrator evolution.

   Only a limited number of suitable calibrators are contained in the
ICRF catalog.  Over 2000 additional calibrators can be found in the
VLBA surveys of calibrations (Beasley \etal~\cite{bea02}; Fomalont
\etal~\cite{fom03})\footnote{http://www.aoc.nrao.edu/vlba/VCS1 and
VCS2}, and most have derived positions which are within 1 mas of the
ICRF grid because they have been observed in one or more of the CDP
24-hour sessions.  Independent searches for calibrator sources, which
are near a desired target source, also can be undertaken using
preliminary observations of km-sized arrays (VLA, WSRT, ATCA) to
search for faint, compact flat-spectrum, candidate sources.  At 1.4
GHz, the field of view of many VLBI arrays are sufficiently large so
that a calibrator can usually be found sufficiently close to the
desired target, and both can then be observed simultaneously.

\begin{figure}
  \qquad \includegraphics[width=6cm]{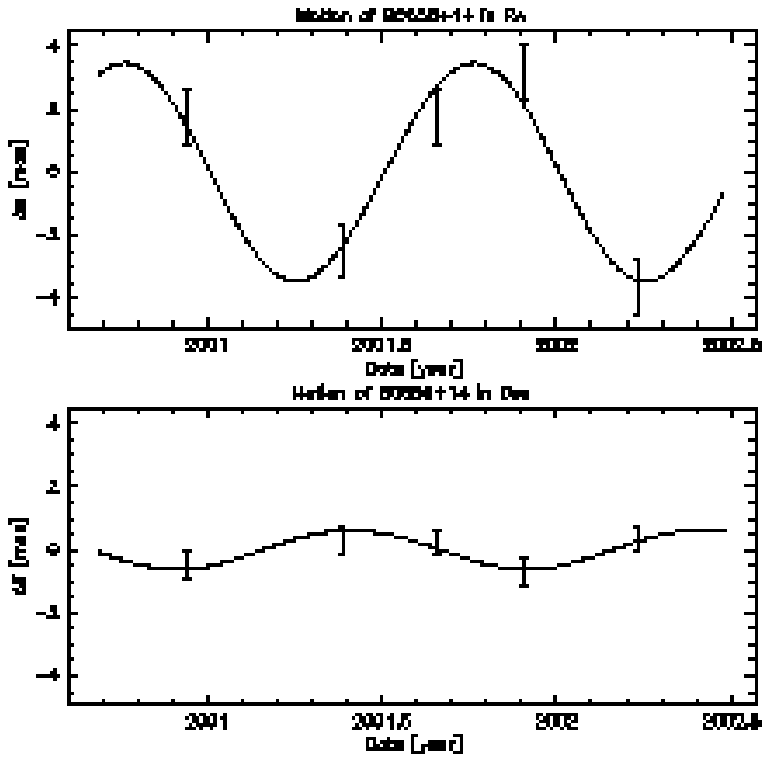} \qquad
  \includegraphics[width=3cm]{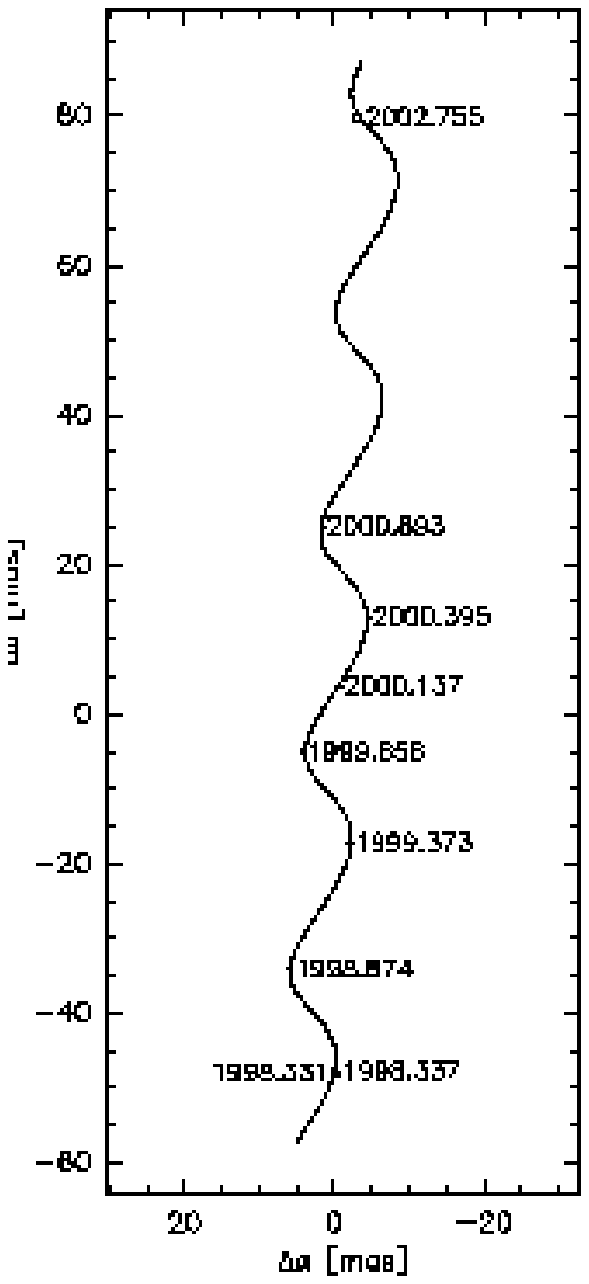} 

  \caption{{\bf(left)} The e/w and n/s parallactic motion of pulsar
  B0656+14 with respect to an in-beam calibrator $16'$ away.  The fit
  is shown by the curve.  {\bf (right)} The motion of pulsar B0950+08
  using a calibrator that is $3^\circ$ away, but using a
  multi-frequency technique for removing the differential ionospheric
  refraction.  The best fit proper motion and parallax are shown by
  the curve.}
\end{figure}

\section {Relative Astrometric Results}

    In this section several phase-referencing astrometric experiments
illustrate the techniques which are currently being used.  The first
example for B0656+14 illustrates a simple phase-referencing (in-beam)
experiment.  The second example for B0950+08 describes a scheme for
removing ionospheric refraction using multi-frequency observations.
The third example for HD8703 illustrates observations over many years
to determine the orbital motion of a resolved radio source, and the
problems associating with calibrator structure changes over time.  All
three of these examples assume that the phase error of the calibrator
is a perfect measure of the phase errors of the target.  The last
example, the measurement of the deflection of a quasar by the jovian
gravitational field, shows that by using more than one calibrator,
relative positional accuracies less than 0.01 mas can be obtained from
a six-hour observation.

\subsection {Simple Phase Referencing}
   The pulsar PSR B0656+14 was observed with the VLBA at 1.67 GHz for
five sessions, separated by about six months each, between epochs
2000.9 to 2002.5 (Briskin \etal~\cite{bri03}).  A calibrator,
J0658+1410 with flux density 35 mJy, was found by preliminary VLA
observations near the pulsar, only $16'$ away, within the reception
area of the VLBA antennas, so that both `in-beam' calibrator and
target were observed simultaneously.  Because of the proximity of the
calibrator and target, the differential delay errors were assumed to
be zero.  The calibrator image was sufficiently point-like; hence, its
structure phase was assumed to be zero and its apparent position
was assumed fixed in the sky since
it is almost certainly an extra-galactic radio source.  Images of the
pulsar were then made from $\triangle\phi^{s-c}$ using the standard
Fourier-techniques.  The pulsar position was determined from the
location of the peak of the pulsar image, with error estimates.  Its
peak flux density was 3.6 mJy and was far too weak to be detected in a
single observation and to derive a group delay.

     Using the images of the pulsar made from each session, the proper
motion and parallax were accurately determined.  The motion associated
with the parallax motion (after removal of the proper motion) is shown
in Fig.~4 (left).  The parallax $\pi=3.47\pm 0.36$ mas, corresponding
to a distance of $291\pm 30$ pc.  The RMS of the fit for each epoch
was about 0.6 mas which is impressive for a relatively weak radio
source.

\subsection{Ionosphere Removal} 

Another experiment, also a measurement of the motion of a pulsar, did
not have the luxury of an available in-beam calibrator.  Using the
VLBA, alternating observations of the pulsar B0950+08 and calibrator
J0946+1017, about $3^\circ$ away, were made every minute over a period
of 7 hours, for 4 sessions between 1998.33 and 1999.85 (Briskin
\etal~\cite{bri00}).  With the relatively large calibrator-target
separation, the ionospheric refraction at 1.6 GHz was expected to be
the dominant source of error.  Since the pulsar is strong, about 64
mJy, the observations of the pulsar and calibrator were made at eight
frequencies, simultaneously, between 1.31 GHz and 1.71 GHz.  The
differential phase error terms in Eq. (\ref{eq2}) can then be written
more generally as
\begin{equation}
 \label{eq3}
   \triangle\phi^{s-c}_{l,m}(t) = \frac{\nu}{c}
       \tau^{s-c}_n(t) + \frac{1}{\nu c} I^{s-c}(t)
\end{equation}
where $\tau^{s-c}_n$ contains all of the non-dispersive delay terms
(independent of frequency), and $I^{s-c}_i$ is the expected
ionospheric delay component---between the two sources.  By fitting the
differential phase observed over the eight frequencies to a function
of the form $A(t)\nu + B(t)\nu^{-1}$, both delay terms can be
determine, and the ionospheric contribution removed from the observed
differential phase.  For telescopes more than 1000 km apart, the
ionospheric-induced position shift can be as large as 10 mas per
degree of source-calibrator separation, changing significantly within
a time-scale of a few minutes; however a shift of 1 mas per degree
over five to ten minutes of time is more typical for the source
position jitter.  Pulsar images, made from the phase before the
ionosphere correction, were severely distorted.

     The pulsar position was then obtained from the ionospheric-free
images, and the position of B0950+08 determined in the same manner as
for B0656+14.  The structure phase of the calibrator J0946+1017 was
determined at each observation epoch and removed from the differential
phase.  The calibrator appearance did not change significantly, hence
it was assumed to be stationary in the sky.  The derived motion of the
pulsar is shown in Fig.~4 (right).  The proper motion is
$\mu_\alpha=-1.6\pm 0.4$ mas yr$^{-1}$, $\mu_\delta=29.5\pm 0.5$ mas
yr$^{-1}$, and $\pi=3.6\pm 0.3$ mas.

\subsection {Complex Motion and Calibrator Stability}

    The Gravity Probe B (GP-B) mission, developed by NASA and Stanford
University, will fly four precision gyroscopes in earth orbit in order
to measure two general relativistic precessions, a geodetic effect and
a frame-dragging effect (Turneaure \etal~\cite{tur89}).  These
precessions will be measured with respect to a suitably bright star
whose position must be known to an accuracy of $<0.5$ mas.  Thus, the
goal of the associated radio observations was to find a suitable
bright star with sufficient radio emission, and to determine its
position and motion with the necessary accuracy.  After exploratory
observations, the RS Can Ven binary-star system, HR8703, was chosen.
The radio emission comes from a main-sequence F-G star and the
companions is a fainter K star with a known period of 24.65 days.
This project illustrates some of the problems associated with both the
target and the calibrator with variable source structure.

\begin{figure}
  \qquad \includegraphics[width=5cm]{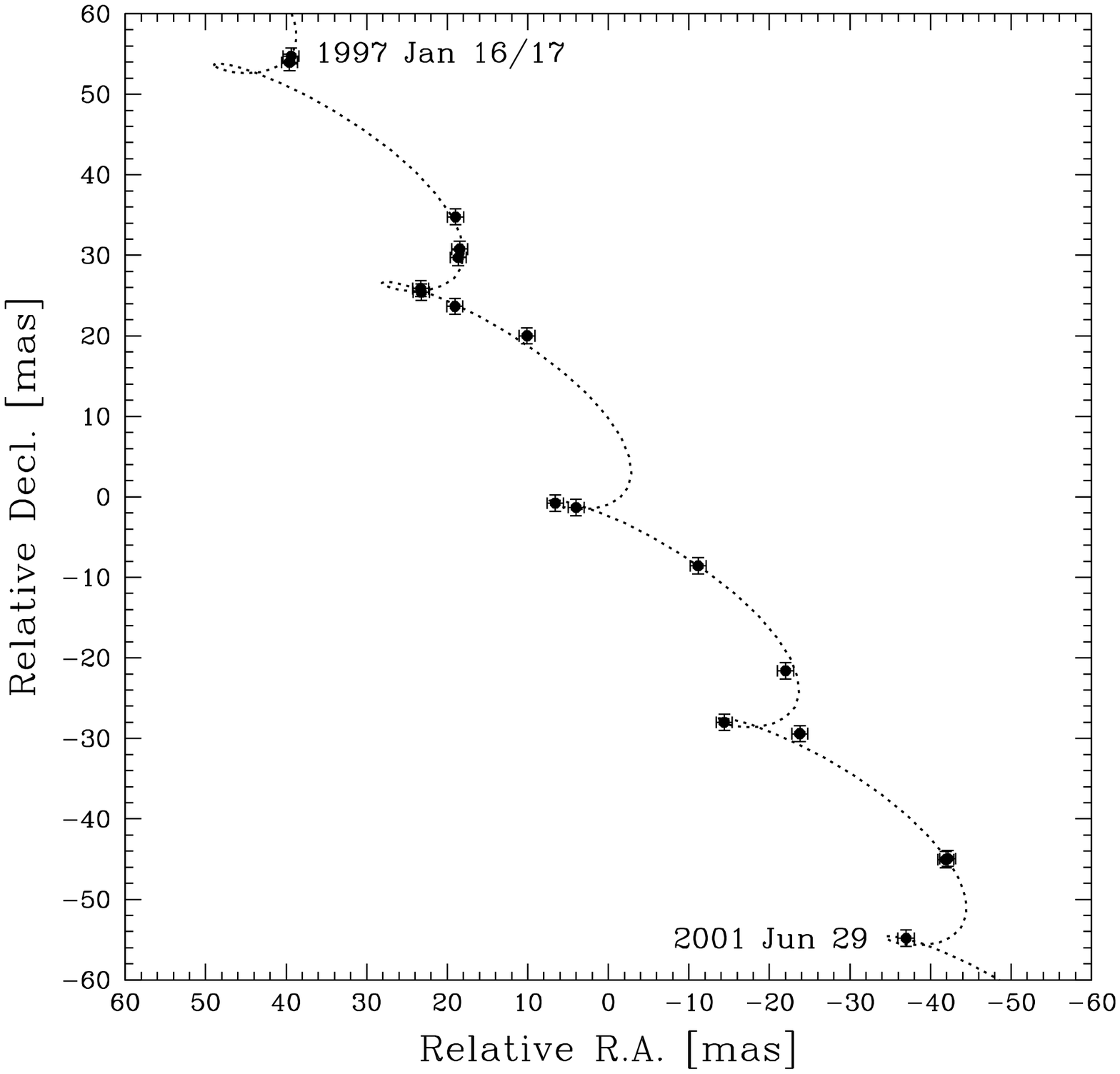} \qquad
  \includegraphics[width=5cm]{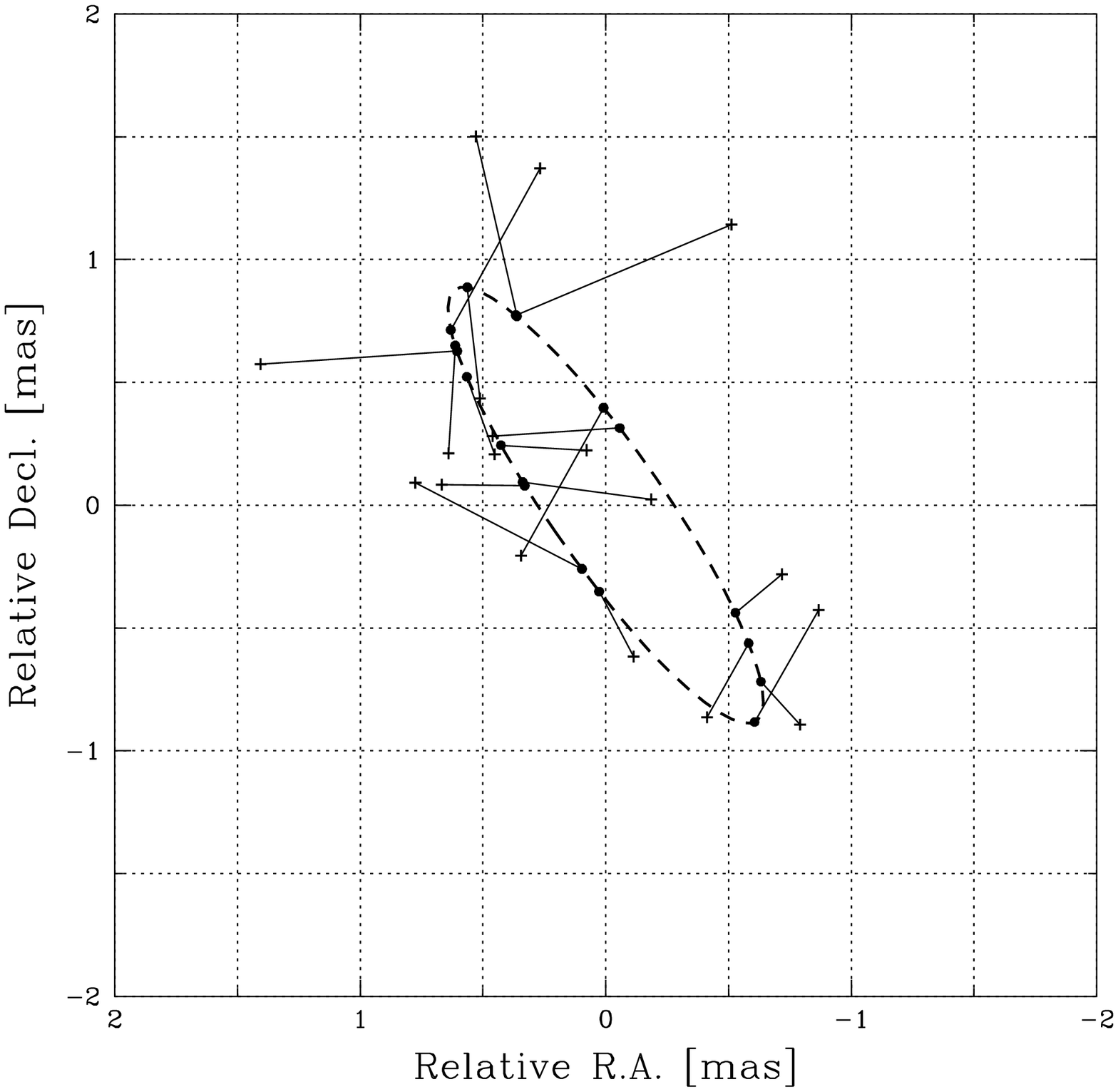} 

   \caption{{\bf (left)} The motion of HR8703 from phase referencing
VLBI observations.  The dashed line shows the best fit of the parallax
and proper motion. {\bf (right)} The dashed line shows the best fit
orbit motion associated with the binary star HR8703 residual motion
after removing the parallax and proper motion, shown on the left.  The
plotted crosses show the measured residual position, and the dots show
the expected location on the orbit.}
\end{figure}

    The radio position of HR8703 has now been measured from more than
20 observation sessions since early 1997, and is still continuing.
The VLBA, the VLA and the three NASA Deep-space Network telescopes, at
8.4 GHz, had sufficient sensitivity that accurate group delays could
be obtained for the star when it was stronger than 5 mJy, which was
most of the time.  The primary calibrator used was 3C454.3 (2251+158),
about $1^\circ$ from the star.  A second calibrator source, B2250+194,
about $3^\circ$ away, was also used, as a control object to determine
the accuracy of the observations.  There are two groups performing the
data reduction.  One group is using the differential phase between
HR8703 and 3C454.3, and the preliminary result reported here is from
this group (Ransom (\cite{ran03}).  Another group is analyzing the
experiment using the group delay.

   Using phase-referencing of HD8703 with 3C454.3 as the calibrator,
good quality images of the binary system were made for each session.
For some of the sessions, the radio emission showed two radio
components separated by 1 mas, in some cases only one component was
observed determined.  The best fit to the parallax and proper motion
(Fig.~5 (left)), was obtained by using the peak position of the radio
emission when one component was detected, and using the geometric
center when two components were detected. (This interpretation is
consistent with the radio emission mechanism of fron RS Can Ven
stars.)  Also, the residual motion of the star, after removing the
parallax and proper motion fit, showed a repeating pattern of 24.65
day period and, hence, this motion was associated with the binary
orbit of the radio emission (Fig.~5 (right).  The estimated error for
the proper motion, parallax and orbital elements is about 0.1 mas.

     The positions of HD8703 are with respect to that of 3C454.3.
However comparison of the relative positions of 3C454.3 (see Fig.~4
(right, bottom)) and B2250+194 over the experiment period show
relative motion between them in the order of 0.3 mas with a time scale
of one year.  Since both calibrators have extended, evolving
structure, it is difficult to determine which source contributes to
the apparent motion, and some of this shift may appear in the
positions derived for HD8703.  Taken over the five years of
observations, these position errors lead to an proper motion
uncertainty of $\sim 0.1$ mas yr$^{-1}$, which is still within the
limited needed for GP-B mission.  However, a more detailed analysis of
the calibrator changes and other phase error terms in Eq.~(\ref{eq2})
is currently underway.

\subsection {Multi-source Calibration Schemes and the Jupiter Deflection Experiment}

    On September 8, 2002 when Jupiter passed within $3.7'$ of a
background quasar J0842+1835, its gravitational field produced an
apparent change (deflection) in the position of the quasar.  At
closest approach, according to GR, the
deflection has two major components: a radial deflection of 1.19 mas,
and a smaller deflection in the direction of motion of Jupiter of
0.051 mas\footnote{A similar experiment in 1988, when Jupiter passed a
close to different quasar, was accurate enough to detect the radial
deflection (Treuhaft \& Lowe \cite{tre91}).}.  This smaller deflection
component is caused by the gravitational aberrational produced by the
relative motion of Jupiter and the earth, and is related to the speed of
propagation of gravity (Kopeikin \cite{kop01}; Frittelli
\cite{fri03}).  Thus, by measuring this aberrational component to an
accuracy less than 0.01 mas, an estimate of the speed of gravity could
be obtained.  However, this precision was about a factor of three to
five better than had been previously obtained (see previous
experiments as good examples).  Such a gain in positional accuracy
could only be obtained by determining the phase error terms associated
with Eq. (\ref{eq2}) more accurately.

\begin{figure}
  \includegraphics[width=6.0cm]{fig6a.ps}
  \includegraphics[width=6.0cm]{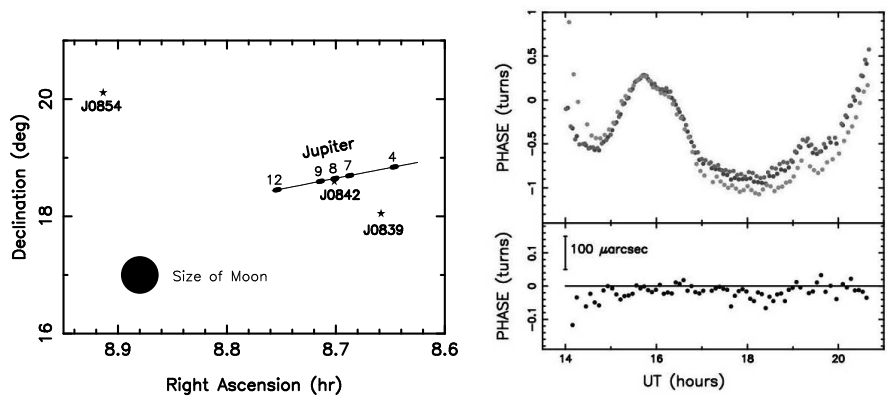} \caption{{\bf
  (left)} The Experimental Configuration for the Jupiter Bending
  Experiment: The observation periods are shown by the heavy portion
  of the Jupiter path, and the position of the three sources are
  nearly colinear. {\bf (right)} The measured phase of the three
  sources on 2003 September 9 for the baseline between Mauna Kea, HI
  and Owens Valley, CA.  Each source point corresponds to one minute
  of data, with a point observed every 4.5 min.  The systematic phase
  difference between the sources after UT=17h is clearly shown.  The
  phases separate with the source J0839, J0842 and J0854 phases with
  descending phase at any time.  The differences are consistent with a
  phase gradient in the sky covering the three sources.}
\end{figure}

    Although phase referencing between a calibrator and target removes
much of the temporal dependence of the phase error, any coherent
angular dependence in the sky will produce systematic phase errors
between the target and calibrator, and produce distorted images and
systematic errors in the position of the peak of the emission.  This
error can be diminished by choosing a calibrator closer to the target
(if you can find one), or by using {\it more than one calibrator} to
determine the angular dependence of the phase error.  Preliminary VLBA
test observations in 2001 suggested that such a multi-calibrator
scheme was effective.  Additional details on multi-source calibrations
are given by Fomalont (\cite{fom03c}).
    
    The design of the Jupiter deflection experiment is shown in Fig.~6
(left).  The observations were made at 8.4 GHz with the VLBA and
Effelsberg, Germany telescope.  In order to determine the phase errors
toward the target source J0842, two calibrators, J0839 and J0854 on
opposite sides of the target, were observed.  Each source was observed
for 1.5 min in turn, and the trio of scans were repeated every 4.5 min
for ten hours over the day.  Although the deflection of J0842 was
strongest on 2002 September 8, identical observations on five days
(September 4, 7, 8, 9, 12) were made in order to obtain sufficient
redundancy to estimate realistic positional errors of the deflection.
More details concerning the design and analysis of this experiment
are given by Fomalont \& Kopeikin (\cite{fom03b}).

Fig.~6 (right, top) shows the measured phase for each of the three
sources for a typical observing day and baseline.  The overall
temporal behavior of the phases, dominated by the clock and gross
troposphere delay errors, are similar for the three sources.  However,
after UT=17h the phases for the three sources show displacements
which are consistent with a simple phase gradient error covering the
three sources: the displacement for J0839 from J0842 is in the
opposite sense and about 25\% of that between J0854 and J0842, as
expected from their relative positions in the sky.  Thus, the
interpolation of the phase of J0839 and J0854 (weighted by their
inverse distance from J0842) gives a very good estimate of the phase
associated with J0842\footnote{In general three calibrators are needed
to determine the phase at the position of the target; two calibrators
are sufficient if they are colinear with the target.  The relative
weighting of the calibrators depend on their distance and orientation
with respect to the target.  Two or even one calibrator with a strong
target may be sufficient to determine the phase gradient if
assumptions are made concerning the phase gradient, for example if it
is elevation dependent}.

The corrected phase of J0842, $\Phi^{J0842}$, after the two source
calibration, is shown in Fig.~6 (right, bottom).  The phase scatter is
small with a RMS less than 0.05 cycle, or about 0.040 mas.  The slight
offset from zero phase is associated with the sum of the position
offsets from the a~priori values, as seen in the expression for
$\Phi^{J0842}$,
\begin{eqnarray}
\label{eq4}
\Phi^{J0842}(t) = \phi^{J0842}(t) - 
      0.80~\phi^{J0842}(t)-0.20~\phi^{J0854}(t)   \\
   =\psi^{J0842} + \frac{\nu}{c}\Bigl[ \big({\bf R_l}-{\bf R_m}\big)
   {\bf \cdot}
   \big(\triangle{\bf K}(t)^{J0842}-0.8~\triangle{\bf K}^{J0839}
   -0.2~\triangle{\bf K}^{J0854}\Bigr] \nonumber \\
   + \hbox{Terms of 2$^{nd}$ order in time and angle} \nonumber
\end{eqnarray}
where the 0.80 and 0.20 are the calibrator weighting factors for this
particular configuration.

The position of the J0842 can be
obtained by Fourier inversion of the corrected phase and measured
visibility amplitude, or by a least-square fit of the phases if the
source is sufficiently strong.  The resultant displacement of the
radio from the image center is a measure of the linear combination of
offset positions (from the a~priori positions), shown in
Eq. (\ref{eq4}).  Assuming that the calibrator positions have not
varied, any changes in the target position with time reflect a change
in the target position.

\begin{figure}
  \includegraphics[width=5.0cm]{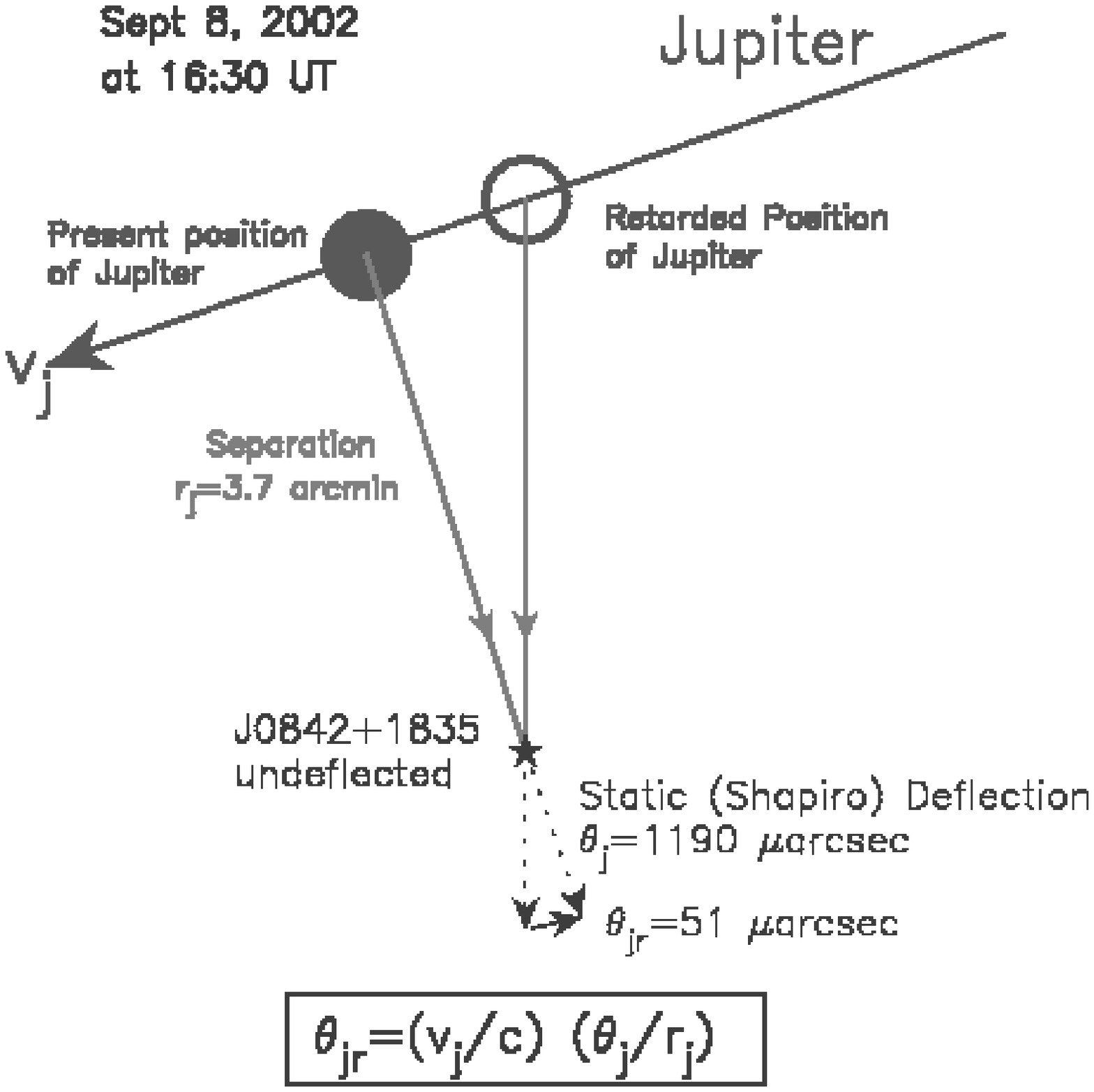} \qquad
  \includegraphics[width=6.0cm]{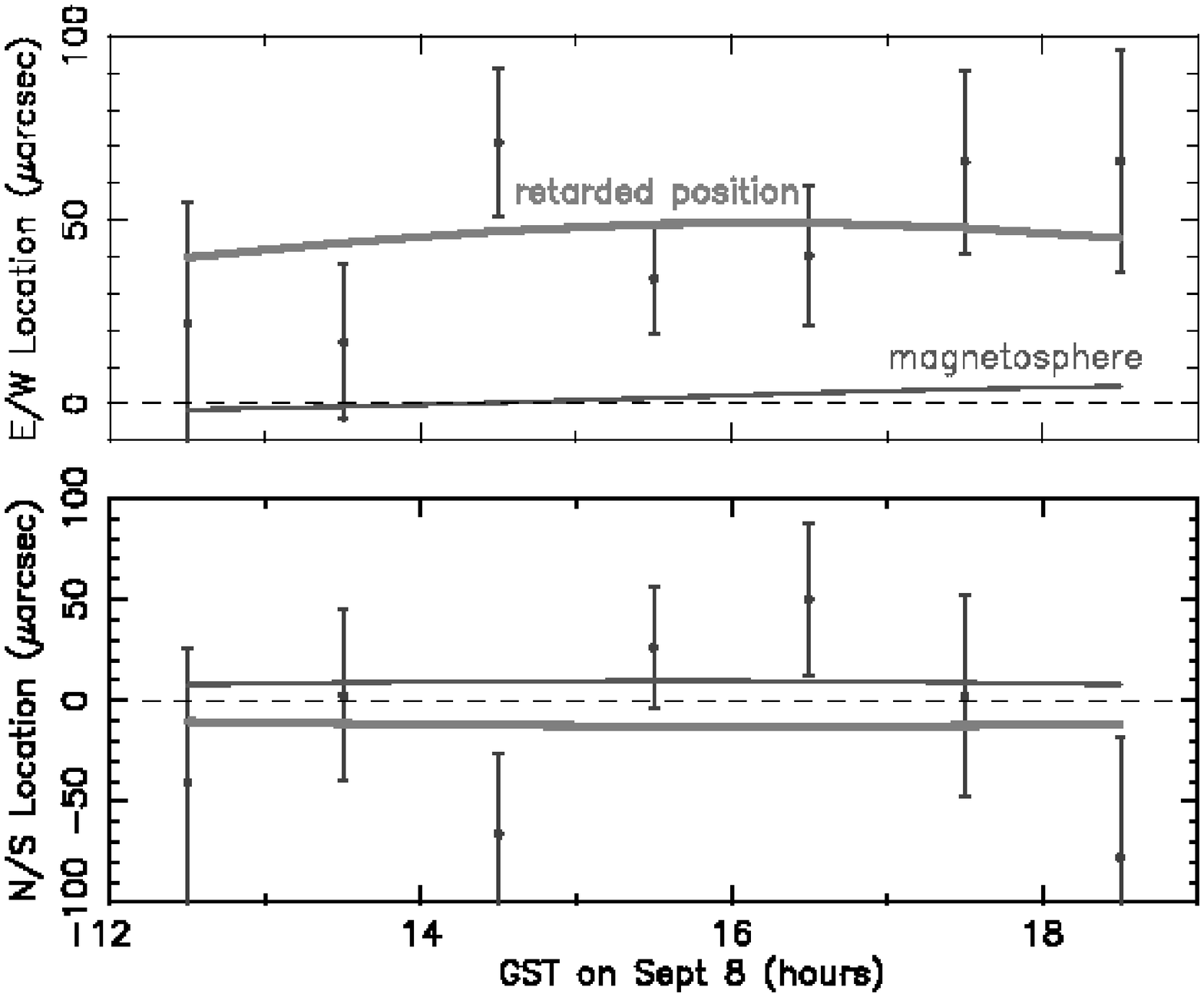} 

   \caption{{\bf (left)} The deflection of the quasar position from
Jupiter is directed from the retarded position of Jupiter.  The
retarded deflection predicted by GR is 0.051 mas in the direction of
Jupiter's motion. {\bf (right)} The measured relative position of the
quasar on 2002 September 8 minus the measured relative position
observed on the other four observing days.  The position change on
September 8 predicted by GR is shown by the curve labeled the retarded
position.  An estimate of the deflection caused by the Jovian
magnetosphere is also labeled.}
\end{figure}

   The goal of the September 2002 observation was to determine the
speed of propagation of gravity, $c_g$, from the determination of the
aberrational part of the quasar deflection, as outlined in Fig.~7
(left).  For $c_g=c$, the expected deflection in the direction of
Jupiter's motion is 0.051 mas at closest approach.  More generally,
the deflection is proportional to $c_g^{-1}$.  A graphical display of
the experimental determination of this deflection term is given in
Fig.~7 (right) which shows the measured position of J0842 every hour
on September 8, relative to the other four observing days.  These
positions were determined from images of the residual phases and the
observed visibility amplitudes.  The measured aberrational deflection,
averaged over September 8, was $0.050\pm 0.009$ mas, which gives a
deflection that is $0.98\pm 0.19$ times that predicted by GR.  The
error estimates are consistent with the position of the target on the
four off-Jupiter days when no change of position is expected.  When
converted into the speed of propagation of gravity, the result is $c_g
= (1.06\pm 0.21)$c.  A similar analysis on the data using only one
calibrator, J0839, gave a positional error of 0.027 mas compared with
0.009 mas achieved using a second calibrator.

\section {Summary}

    After a discussion of inertial frames and reference systems, the
methods for determining the position of radio sources were discussed.
First, all-sky observations used for the ICRF were used to determine
absolute positions of radio sources which formed the definition of the
celestial inertial reference system now in use.  Secondly, phase
referencing was described using four different experiments to
illustrate several aspects of measuring the motion of radio sources to
accuracies approaching 0.010 mas per session.

\section {Appendix: The Speed of Gravity Controversy}
    Although the experimental result is not in doubt, there is
disagreement over the correct role of gravity in this experiment.
Kopeikin (\cite{kop01}), using the Einstein equations directly,
interprets the retarded deflection component as a measure of the speed
of propagation of gravity (see also Kopeikin \& Fomalont
(\cite{kop04})).  Will (\cite{wil03}) concludes, however, that the
retarded deflection component is associated with the aberration of
light, not of gravity.  Other interpretations are given by Asada
(\cite{asa02}) and Samuel (\cite{sam03}).

    At the Jenam2003 meeting in Budapest, a lively give-and-take
session took place during the late afternoon of August 28.  There were
about 20-30 participants. The main points of view were presented by
G. Sch\"afer (Institute for Theoretical Physics, Friedrich Schiller
University, Jena, Germany), Sergei Kopeikin (University of
Missouri-Columbia, MO, USA), and Gabor Lanyi (Jet Propulsion
Laboratories, Pasadena, CA, USA)

Sch\"afer supported the point of view of Asada and Will that the
experiment measured the speed of radio waves used in observations. The
main arguments were:
\begin{enumerate}

\item gravity and light move along null geodesics and their speed can
not be disentangle in the experiment;

\item the speed of gravity is associated with accelerated motion of
light-ray deflecting bodies and thus, only with gravity waves;

\item the propagation of gravity effect shows up only in terms of order of
$(v/c)^4$ in the metric tensor but the experiment is sensitive to
$(v/c)^3$ terms only.
\end{enumerate}

Kopeikin interpretation is that that the experiment measured the speed
of propagation of gravity in the near zone of the solar
system.  In answer to  Sch\"afer's points, Kopeikin pointed out that:
\begin{enumerate}

\item The relativistic deflection of light by Jupiter depends on the
Euclidean product of two null vectors at the point of
observation---the vector of light propagation from the quasar and that
of gravity propagation from Jupiter.  Because the quasar and Jupiter
are located at different positions on the sky, the gravity propagation
direction can be separated from the light propagation direction and,
thus, the speed of gravity could be measured.

\item There are different types of gravitational fields
according to algebraic classification B. Y. Petrov. Only
gravitational fields of type N (plane gravitational waves) are
generated by the accelerated motion of bodies, and they exist only in
the radiative zone of the solar system. Other types of gravitational
fields are more general and can be associated with the velocity of the
bodies. They dominate in the near zone of the solar system where the
jovian deflection experiment has been done.  Nevertheless,
gravitational fields of the other Petrov's types also propagate with
finite speed which can be measured. 

\item It is true that propagation of gravity effect shows up only in
terms of order of $(v/c)^4$ in the metric tensor. But the metric
tensor is not directly observable quantity in the jovian deflection
experiment because it depends on specific choice of
coordinates. Therefore, any reference to the metric tensor in
connection with measured gravitational effects is irrelevant.  What is
observed is the phase of radio waves coming from the quasar which is
perturbed by gravitational field of Jupiter from its retarded position,
in accordance with Einstein's equation predicting that gravity travels
with finite speed equal numerically to the speed of light in vacuum.
This relativistic effect of the retardation of gravity in the
electromagnetic phase is gauge-independent and shows up already in
terms of order of $(v/c)^3$ which we have observed.

\end{enumerate}
\section {Acknowledgments}
    The sections on reference frames and the ICRF were significantly
improved from the suggestions and corrections by Chris Jacobs.  I also
thank Chopo Ma for his contributions, and Ryan Ranson for the
preliminary HR8703 results.


\end{document}